# Neutron diffraction evidence of the 3-dimensional structure of $Ba_2MnTeO_6$ and misidentification of the triangular layers within the face-centred cubic lattice.

Otto H. J. Mustonen,[1] Charlotte E. Pughe,[1] Helen C. Walker,[2] Heather M. Mutch,[1] Gavin B. G. Stenning,[2] Fiona C. Coomer,[3] Edmund J. Cussen[1*]

[1]Department of Material Science and Engineering, University of Sheffield, Mappin Street, Sheffield S1 3JD, United Kingdom

[2]ISIS Pulsed Neutron and Muon Source, STFC Rutherford Appleton Laboratory, Harwell Campus, Didcot OX11 0QX, United Kingdom

[3]Johnson Matthey Battery Materials, Blount's Court, Sonning Common, Reading RG4 9NH United Kingdom

* Correspondence to e.j.cussen@sheffield.ac.uk

$Ba_2MnTeO_6$ was first characterised using X-ray diffraction and reported to show a small distortion[1] from the idealised cubic perovskite which displays face-centred cubic arrangement of $Mn^{2+}$. A recent report has asserted that this leads to a layered configuration of $Mn^{2+}$ that serves as an example of a triangular lattice, i.e. a 2D structure containing discreet layers.[2] Here we show how neutron scattering gives great confidence in establishing the crystal structure being an undistorted cubic phase and how this can be mis-assigned as a triangular layered structure. This has profound implications for the understanding of the magnetic properties of the system.

Magnetically frustrated systems are readily visualised by considering antiferromagnetic coupling between magnetic centres on the points of an equilateral triangle. Such a configuration can be realised physically in a number of well-studied structures giving rise to kagome, pyrochlore, triangular, and face-centred cubic lattices.[3] The interpretation of the resultant magnetic properties relies on a precise understanding of the geometry of the crystal lattice which underpins and determines the spatial distribution of magnetic ions. $Ba_2MnTeO_6$ contains high-spin ($d^5$) $Mn^{2+}$ embedded in an otherwise diamagnetic crystal structure and the absence of orbital angular momentum in this L=0 ion makes it an excellent candidate to study the impact of lattice geometry on magnetic interactions.

The correct identification of the crystal structure of $Ba_2MnTeO_6$ is of central importance in interpreting the magnetic properties as the magnetic interactions are defined by the geometry of the crystal structure. The magnetic interactions in $Ba_2MnTeO_6$ have been interpreted[2] using the assumption that it is a model 2-dimensional triangular lattice. Such triangular lattices are of wide interest, and examples lead to a variety of novel magnetic states.[4] In this instance, the structure of $Ba_2MnTeO_6$ is actually an isotropic 3-dimensional structure. Here we expand our previous report[5] to explain the unambiguous evidence from high resolution neutron powder diffraction data and clarify the relationship between the 3D cubic structure and the 2D triangular lattices that have been reported.

Distortions in the perovskite structure commonly arise from displacements of the oxide anions that break the $Fm\bar{3}m$ symmetry of the cubic cation-ordered double perovskite aristotype.[6] This can lead to ambiguities in perovskite structure determination where the weak X-ray scattering from oxide anions is largely masked by the presence of dominant scatterers such as $Ba^{2+}$ and $Te^{6+}$ that undergo negligible displacement in response to the reduction of symmetry.

This point can be illustrated by X-ray diffraction patterns calculated from the recent report[2] of $R\bar{3}m$ $Ba_2MnTeO_6$ and compared with the $Fm\bar{3}m$ structure[5] as shown in Figure 1. These data illustrate that laboratory X-ray diffraction data provide no meaningful differentiation between these two structures. The crystallographic convention is that the higher symmetry is used to describe the structure, in this case $Fm\bar{3}m$.

Any ambiguities in analysis of the X-ray profile can be resolved using neutron scattering where the sensitivity to oxide is much greater.[7] High quality neutron diffraction data were collected using the GEM diffractometer[8] and Figure 1 compares the recently reported observed neutron diffraction profile[5] to diffraction profiles calculated using either the $Fm\bar{3}m$ perovskite or the proposed $R\bar{3}m$ structure. Due to the negligible distortion from metric cubic symmetry of the rhombohedral structure both patterns have apparently similar distribution of permitted Bragg reflection positions. The greater sensitivity of the incident neutron wave to the oxygen atoms means that even small displacements of oxide from the special positions of the $Fm\bar{3}m$ structure have a significant impact on the diffracted neutron profile. The simulated pattern for the $R\bar{3}m$ structure shows that several Bragg peaks should have considerably enhanced intensity that is evidently not present in the observed pattern. Thus we conclude that there is no evidence of structural distortion from cubic symmetry in $Ba_2MnTeO_6$ and that the $Fm\bar{3}m$ space group can be assigned with utmost confidence.

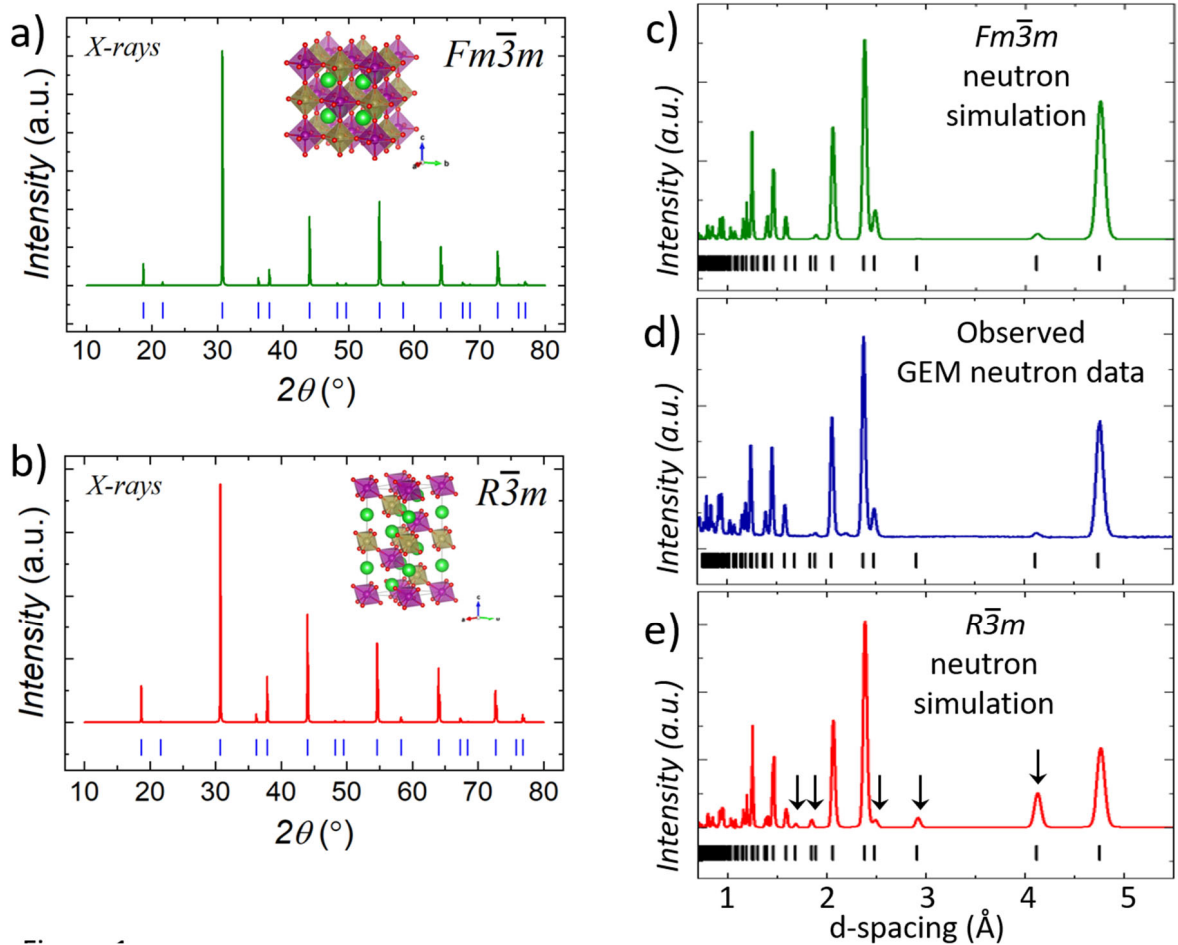

Figure 1 Calculated X-ray diffraction profiles for a) $Fm\bar{3}m$ and b) $R\bar{3}m$ structures using published parameters. The data for $Fm\bar{3}m$ were taken from 100 K structure solution[5]. The displacement parameters For the $R\bar{3}m$ model were not reported[1] and values for $U_{iso}$ were taken as 0.001 Å$^2$ for Ba/Mn/Te and 0.002 Å$^2$ for O as identified in the $Fm\bar{3}m$ solution. There are no large differences between these X-ray diffraction profiles. c) The simulated neutron diffraction pattern from the $Fm\bar{3}m$ solution shows an excellent agreement with the observed neutron diffraction profile d), but shows very large intensity mismatches indicated by arrows in e) compared to the simulated neutron diffraction profile from the distorted perovskite described by the $R\bar{3}m$ structure.

Given the similarity of the X-ray diffraction profiles of the $Fm\bar{3}m$ and $R\bar{3}m$ structures it was unexpected to see that the structure of $Ba_2MnTeO_6$ was being considered as an example of a 2D triangular lattice. The arrangement of Mn and Te cations in the $Fm\bar{3}m$ structure is compared to that reported as a 2D triangular lattice in Figure 2.

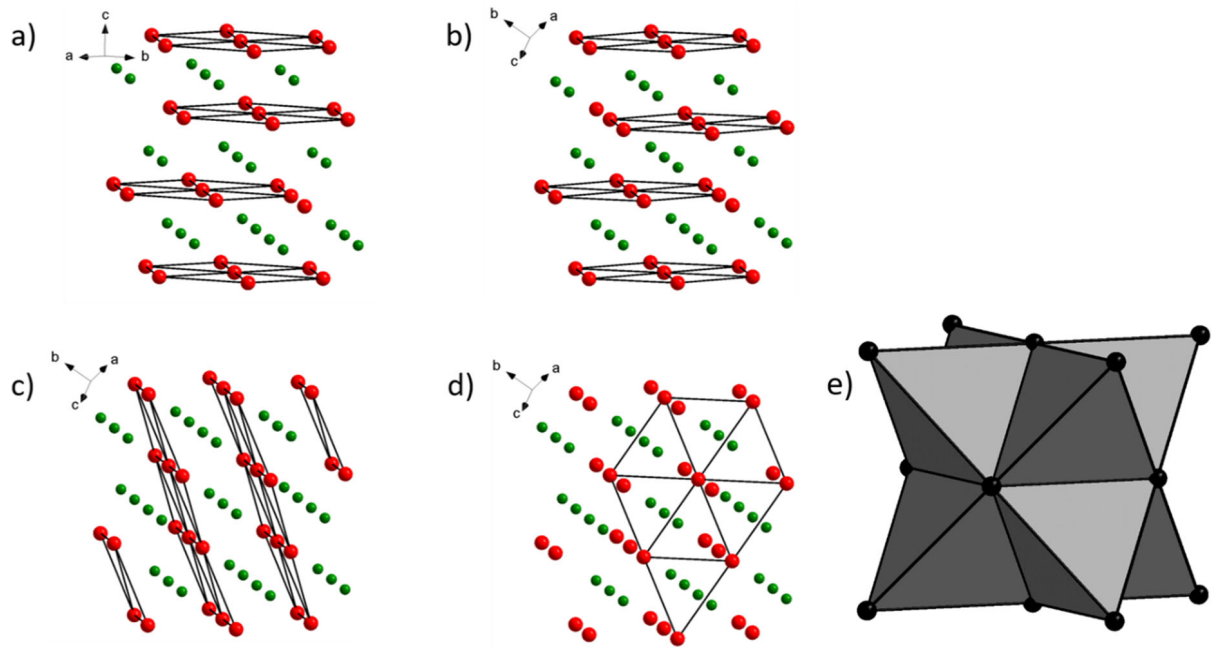

Figure 2 The arrangement of Mn (red spheres) and Te (green spheres) in the crystal structure of $Ba_2MnTeO_6$ derived from a) the $R\bar{3}m$ model, with lines drawn to reproduce the illustration of layers[2] of Mn cations.  b) The $Fm\bar{3}m$ structure derived from neutron diffraction data uses the same projection and scale as (a) to display and selectively links some of the Mn cations that are separated by a distance of 5.8 Å.  This reproduces the suggested 2D layered structure in (a).  Figures (c) and (d) makes connections between other layers of Mn cations that are symmetry equivalent to the linkages shows in (b). This shows that the apparent 2D triangular lattice is an artefact arising from viewing only a single projection, and ignoring the 3D connectivity. The 3D arrangement of edge-sharing equilaterally triangles is a well-established feature[3] of double perovskite structure as shown in (e).

In the $R\bar{3}m$ structure, the apparently layered 2D structure is generated by connecting each Mn cation to six neighbouring ions at a distance of 5.816 Å to form a hexagonal net composed of triangles.  However, each Mn ion has another six neighbours at an almost identical distance (5.817 Å) that connect into adjacent 'layers'. In the $Fm\bar{3}m$ structure these two distances become symmetry equivalent. The apparent layering within the $R\bar{3}m$ structure can be achieved by selectively drawing half of the linkages between neighbouring Mn ions and omitting to draw the other half. Once the connections are drawn to link all nearest Mn-Mn pairs it is clear that the 3D structure is composed of edge-sharing equilateral triangles that is a familiar feature of the face-centred cubic lattice.  Understanding the difference between these lattice types is of fundamental importance to any investigation of magnetic properties.  The three-dimensional connectivity arising from the cubic symmetry in $Ba_2MnTeO_6$ negate any analysis based on intra- and inter-layer coupling between arbitrarily selected layers.

**Acknowledgements:** This work was funded by the Leverhulme Trust Research Project Grant RPG-2017-109. The authors are grateful to the Science and Technology Facilities Council for the beamtime allocated at ISIS and to Dr Ivan da Silva for assisting with this data collection and to Dr Alex Gibbs for assistance and helpful discussions.

**Competing Interests:** The authors declare that they have no competing interests relating to this work.

**Author contributions statement:** OM, CP, HM contributed to all aspects of the material synthesis, data collection and analysis.  HW and GS assisted with the neutron scattering analysis and FC and EC planned the project.  The manuscript was drafted by CP, OM and EC with input from all authors.

**Correspondence** and requests for materials should be addressed to E.J.C. e.j.cussen@sheffield.ac.uk